\documentclass[journal]{IEEEtran}
\usepackage{cite}
\usepackage{amsfonts}
\ifCLASSINFOpdf
\usepackage[pdftex]{graphicx}
\else
\fi
\begin{document}
%
\title{A Novel Tunable Controller for Grid Forming Converters towards Critical Services Application}
%
%
%

\author{Yangyadatta Tripathy,~\IEEEmembership{Student Member,~IEEE,} Barjeev Tyagi,~\IEEEmembership{Member,~IEEE,}
\thanks{Yangyadatta~Tripathy is with the Department
of Electrical Engineering, Indian Institute of Technology, Roorkee,
IND, 247667 INDIA e-mail: (y\_tripathy@ee.iitr.ac.in).}}

%
%

\markboth{Journal of \LaTeX\ Class Files,~Vol.~14, No.~8, August~2015}%
{Shell \MakeLowercase{\textit{et al.}}: Bare Demo of IEEEtran.cls for IEEE Journals}
%


\maketitle

\begin{abstract}
This paper demonstrates the key features of a control system applicable to inverter-based resources (IBR), which is based on grid-forming technology. Such resources are classified as grid-forming or grid-following converters based on the type of output with or without grid connection. With rapid growth in the energy sector to adopt carbon-free generation, Grid Forming Converter (GFC) seems suitable for power provision to remote or islanded operation of converters. A fully-fledged bulk power grid based on GFC requires complex control implementation with suitable tuning of its parameters. In this article a broader analysis of synchronous machine and such type of converter is discussed and designed in the MATLAB 2024 environment with its control technique is studied for a closed-loop system under contingencies. A proposed control scheme is developed to understand the frequency minimization problem and the minimization problem is solved using GAMS programming tool. The primary objective function is found to be suitable for minimization of frequency deviation using a mixed control approach. An artificial neural network-based controller is also proposed with Levenberg-Marquardt training algorithm which augments the research by finding suitable optimal reference for GFM converter in the presence of a grid. A long-short-term memory (LSTM) based network is also proposed for the above control and the performance is found to be efficacious.
\end{abstract}


\IEEEpeerreviewmaketitle

\section{Introduction}

As the renewable-based generation in the energy portfolio mix of the country increases, the conventional plants are augmented with converter-based generation technologies like Grid-forming converter (GFC) and Grid-following converter (GFL) based plants. The control of such power-electronic switches for rapid power production, as well as emulating the synchronous dynamics, is very difficult to achieve. The switches face enormous switching stress which is alleviated through appropriate cooling methods. The control centers were equipped with large scale thyristor valves and high-end processors with precise and accurate controls to achieve grid formation \cite{Rosso2021, Unruh2020, Song2022}.\par
The ideal grid-forming converter can be represented as an active voltage source connected with an impedance in series with the source, which forms the grid by providing power to the critical loads. It does not require the phase locked loop (PLL) as its counterpart, i.e., grid following converters (GFL). Grid-following converters are represented as a modular current source with a parallel impedance providing power to the grid. GFL supplies current to the grid by taking grid voltage as reference via PLL \cite{roca2012}. The lack of grid operations in GFL inverters brought system stability and reliability issues and control challenges as GFL cannot provide any inertia and lower the overall inertia. The decrease in overall inertia leads to vulnerability in grid variations, such as weak grid conditions, where short circuit strength is low. In case of Grid forming converters, power systems operations are enhanced, e.g., regulated voltage and frequency in synergy with power grids. The GFM converters provide droop control function, virtual oscillator function, and virtual synchronous generator functions, which provide voltage, frequency, and inertia support to the power grid, respectively. Such support functions allow GFMs to operate with parallel converters. GFM converters have many functions which are well researched, i.e., self-synchronization, seamless mode transfer, as well as black-start. Self-synchronization function utilises the droop control and common DC-link capacitor voltage control, and combines both functions. For a flexible transition between grid-connected mode and islanded mode, a seamless mode transfer function is deployed. In the black-start function,  restoration of power in case of blackouts is done. Islanding detection in GFM is also an active research area which is explored in \cite{bkpani2025} using deep learning architecture which employs long short-term memory-based feature classifier to overcome manual threshold selection in this research.  Table \ref{tab:sota} recommends the state of the art research from the academic community about the challenges and summary of the research contributions.
\begin{figure}[h]
    \centering
    \includegraphics[width=1\linewidth]{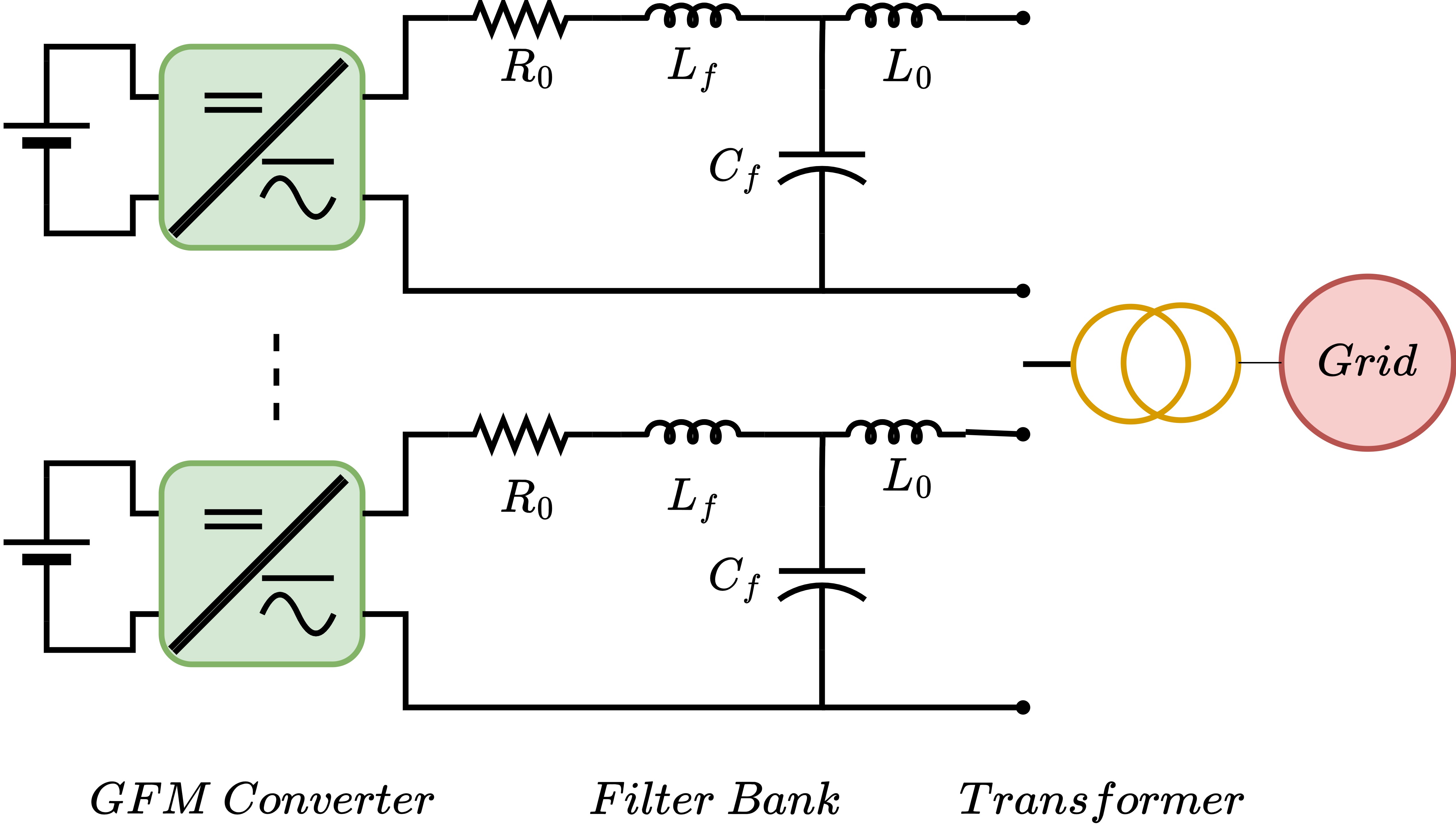}
    \caption{Block diagram of multiple GFM converter in presence of grid}
    \label{fig:enter-gfm}
\end{figure}

This paper reviews the state-of-the-art control methods developed by researchers in the last decade for GFM converters. 

In \cite{Harrison2021} paper assesses the equivalence of Grid Forming (GFM) droop and Virtual Synchronous Machine (VSM) controls, finding that while they are equivalent in steady state, the GFM droop is less damped in dynamic response, and adjusting Voltage Control parameters can improve but not fully achieve equivalence with the VSM. The methodologies used are 
\begin{enumerate}
    \item tuning controls using a transfer function (tf) method to match synchronous machine properties. Deriving transfer functions to compare system properties of GFM droop and VSM controls.
    \item Using time-domain models to simulate control behavior under frequency disturbances.
   \item Conducting a parametric sweep to analyze cascaded control dynamics.
   \item Comparing steady-state and dynamic responses of GFM droop and Synchronverter.
\end{enumerate}
The limitations of the above research is
\begin{enumerate}
    \item Equivalence generally focuses on individual operating conditions, not dynamic properties.
    \item Different inertia implementation leads to significant differences in control responses.
   \item Difference in dynamic response due to cascaded control or non-equivalence between GFM strategies.
   \item Steady-state equivalence does not hold for all values of inertia, damping, and short circuit ratio.
\end{enumerate}

In \cite{YunYU2021}, paper provides an overview of grid-forming control algorithms for wind turbine converters, comparing their characteristics for voltage build-up, synchronization, interaction attenuation, and virtual inertia. The above research is supported with
\begin{enumerate}
    \item Wind turbine converters with grid-forming control can support voltage and frequency in power systems by emulating synchronous machine dynamics.
    \item The study provides a comprehensive overview of state-of-the-art GFM control algorithms and their characteristics. 
    \item Future research needs to focus on mitigating oscillations, improving inertia emulation, and coordinating with energy storage systems. 
\end{enumerate} However, mitigation of oscillations among WTCs, appropriate inertia emulation, and coordination with energy storage systems need to be addressed.
In \cite{Jiang2024} comparison of PLL-based and PLL-less grid-forming converters reveals their similarities in grid-supporting functionalities and differences in dynamic response, synchronization stability, and small-signal stability. A comparative study is conducted between PLL-based and PLL-less grid-forming converters to assess their synchronization methods. Both types of converters have similar grid-supporting functionalities and current-limiting capabilities despite differences in synchronization. The study provides insights into the impact of synchronization loops on dynamic response, synchronization stability, and small-signal stability. However, PLL
based GFM converters should be carefully designed to enhance its stability margin in weak grid operation. One of the drawback of this oepration is by changing PLL gains harmonics can be eliminated yet the system tends to be marginally stable.

In \cite{Papadopoulos2021}, investigation on small-signal interactions between synchronous machines and grid-forming converters, including the possibility of electromechanical oscillatory modes is conducted. The paper confirms the possibility of electromechanical oscillatory modes between conventional synchronous machines and grid-forming converters, as well as between two grid forming converters. The study uses small-signal models and eigenvalue analysis to investigate these interactions. Parametric sweeps were performed to investigate the impact of network and GFM control parameters on controlling these oscillatory modes. Small-signal instability is found to occur from high values $K_p$ of as the electromechanical mode traverses into the unstable region. In the case of $K_p$, this mode can be fully damped but doing so will bring a higher frequency oscillation towards instability.
In \cite{SNag2023} a unified control design for software-defined inverters to operate in either grid-forming or grid-following mode while enforcing grid operational constraints is proposed. The proposed control Unifies control design for software-defined inverters (SDIs) in grid-forming (GFM) and grid-following (GFL) modes. Development of such control schemes based on dynamic state space models is studied for nonlinear nominal output tracking control and constraint-enforcing control. It provides recursive Lyapunov design paradigm for tracking and control laws.
 Section II and III provides modelling of synchronous machine and the control alogrithms for grid forming converter control respectively,
 Section IV starts to describe different control approaches of GFM converters and their applications. Section VI literature introduces the critical services sector.The conclusion and future scope are presented in the last section for different GFM technologies.

\section{Operating Principles of Synchronous Machine and GFM Comparison}
Synchronous Machines (SM) are best known for their voltage forming ability and inertia provision during any grid disturbance. The underlying principle behind SM operation is a magnetized rotor rotates to induce voltage in three phase stator winding. The rotor is excited by a permanent magnet or an external supply. Mathematically, the flux linkage between the rotor and stator is simplified when both rotor and stator variables are in a coordinate system of the speed of the rotor. Otherwise, the inductances were present in the equations and combined in such a way that they varied with the position of the rotor. Such a transformation of the coordinate system was first proposed by Park, which is known as Park's transformation or direct-quadrature-zero (dq0) transformation. An extension to this theory is proposed in \cite{2react}.

\begin{figure}[h]
        \centering
        \includegraphics[width=1\linewidth]{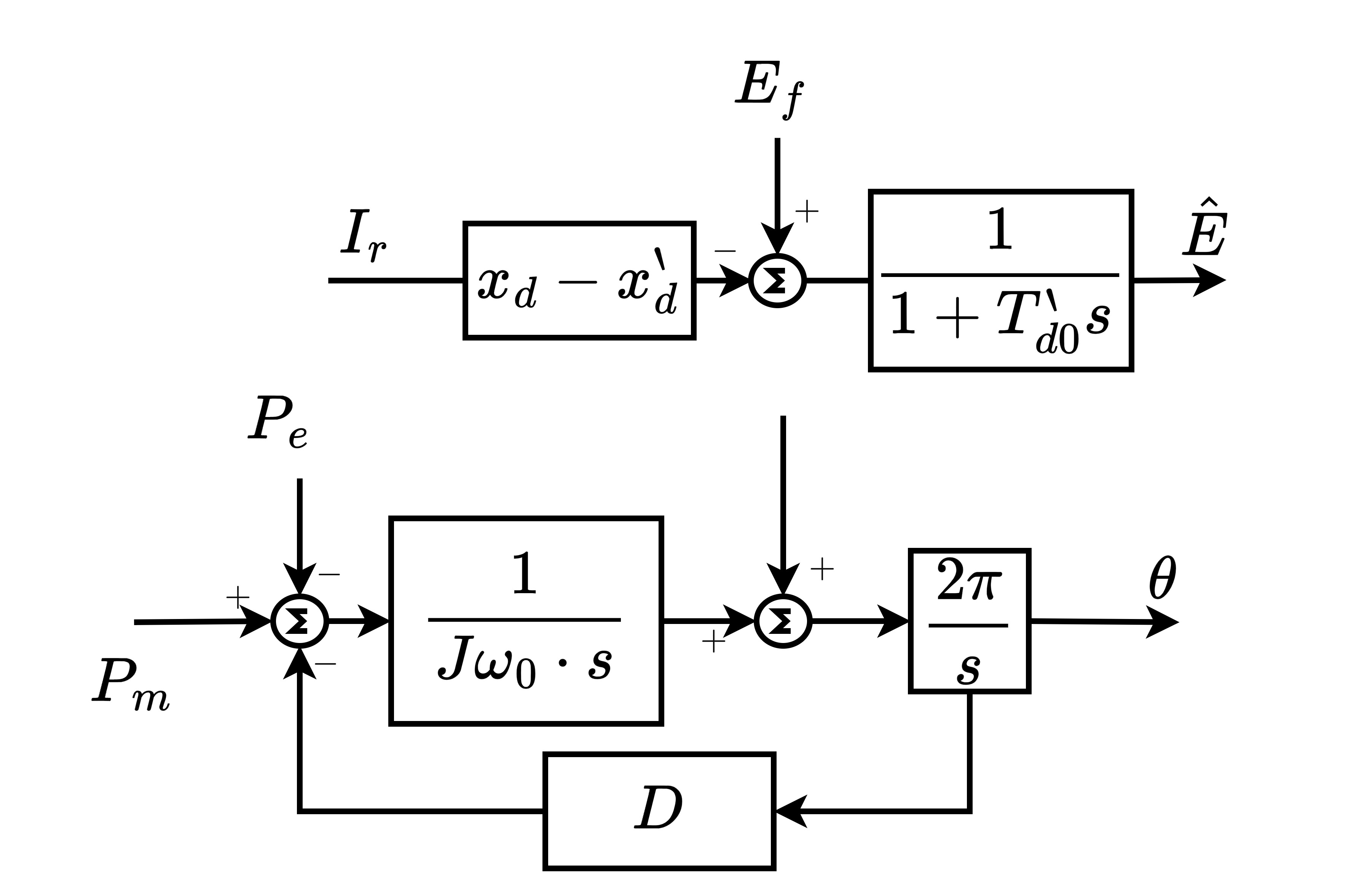}
        \caption{Block diagram of 3rd order synchronous machine model}
        \label{fig:enter-smmodel}
\end{figure}

The existing model of the synchronous machine contains higher-order terms that represent the accuracy of the machine model. A tradeoff between complexity and accuracy is made to obtain lower order models by model order reduction procedures. An example of order reduction is to neglect the armature resistances, the high transient behavior of ammortissers. By incorporating such assumptions, a one-axis third-order model can be obtained. An electromagnetic link between stator and rotor is obtained by the dynamics, which is shown, and another equation represents the swing behavior, which is shown below.
\begin{equation}
    \hat{E}=\frac{E_f}{T_{d0}^`s+1}+\frac{x_d-x_d^`}{T_{d0}^`s+1}I_r
\end{equation}

\begin{equation}
    J\omega\frac{\partial \omega}{\partial t} = P_m-P_e
\end{equation}
where
$\omega = \frac{\partial \theta}{\partial t}$

Equation (1) represents the internal voltage of SM, which consists of two parts, e.g., $E_f$ and reactive current $I_r$, separated by the occurrence of effects by a delay of the open-circuit time constant $T_{d0}^`$. So, in terms of transfer functions, the same can be written as follows.

\begin{equation}
    G(s) = \frac{1}{T_{d0}^`s+1}
\end{equation}
and
\begin{equation}
    x_d(s) = \frac{x_d-x_d^`}{T_{d0}^`s+1}+x_d^`
\end{equation}
Equation (4) can be interpreted as after the first moment of the current step function, only the transient reactance is active. In the meantime, the reactance decays to synchronous value $x_d$.

\begin{figure}[h]
    \centering
    \includegraphics[width=0.75\linewidth]{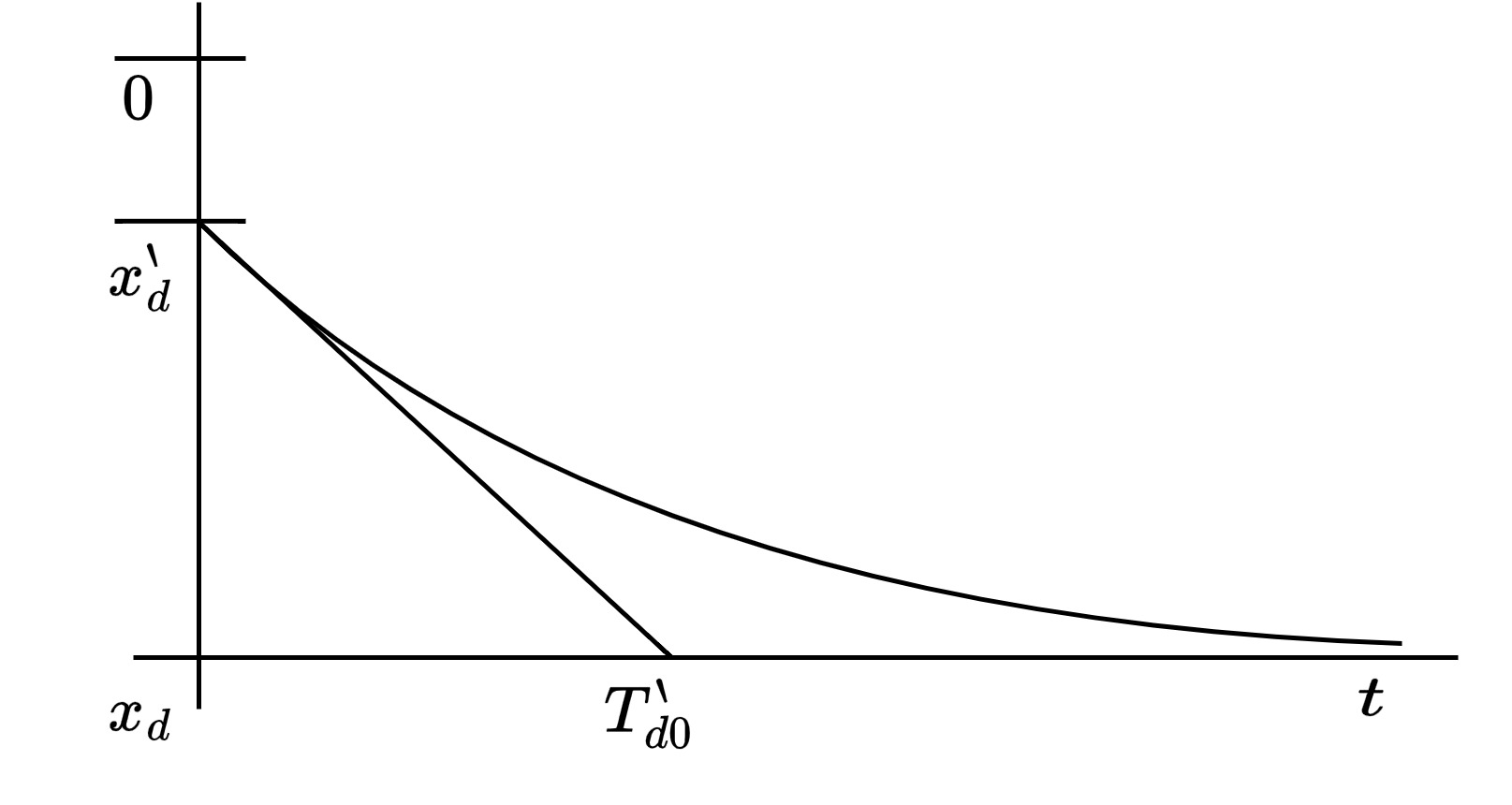}
    \caption{Impedance evolution after step current of synchronous generator}
    \label{fig:enter-imp_evol}
\end{figure}

\section{Grid forming Control methods}
\subsection{Factors Affecting Angle Reference in Grid-Forming Control}

In grid-forming (GFM) control, the voltage angle reference $\theta_{\mathrm{ref}}$ governs the instantaneous phase of the generated voltage waveform. The dynamics of $\theta_{\mathrm{ref}}$ are influenced by multiple physical and operational factors, which are outlined below.

\subsubsection{Active Power–Frequency Droop}
In conventional $P$–$f$ droop control, the instantaneous frequency is adjusted based on active power deviation:
\begin{equation}
    \omega(t) = \omega_{\mathrm{nom}} - K_p \left( P(t) - P_{\mathrm{ref}} \right)
    \label{eq:droop}
\end{equation}
The angle reference evolves by integrating the instantaneous frequency:
\begin{equation}
    \theta_{\mathrm{ref}}(t) = \theta_{\mathrm{ref}}(0) + \int_{0}^{t} \omega(\tau) \, d\tau
    \label{eq:angle_integration}
\end{equation}

\subsubsection{Virtual Synchronous Machine (VSM) Dynamics}
In VSM control, the angle dynamics are determined by the swing equation:
\begin{equation}
    J \frac{d\omega}{dt} = P_{\mathrm{m}} - P_{\mathrm{e}} - D \left( \omega - \omega_{\mathrm{nom}} \right)
    \label{eq:vsm}
\end{equation}
The angle reference is obtained from:
\begin{equation}
    \frac{d\theta_{\mathrm{ref}}}{dt} = \omega(t)
    \label{eq:vsm_angle}
\end{equation}

\subsubsection{Reactive Power–Voltage Droop Coupling}
The $Q$–$V$ droop relation indirectly impacts $\theta_{\mathrm{ref}}$ through network coupling:
\begin{equation}
    V(t) = V_{\mathrm{nom}} - K_q \left( Q(t) - Q_{\mathrm{ref}} \right)
    \label{eq:qv_droop}
\end{equation}
In inductive networks, active power flow is related to phase angle as:
\begin{equation}
    P \approx \frac{V_1 V_2}{X} \sin\delta
    \label{eq:power_angle_relation}
\end{equation}
showing that changes in reactive power may indirectly influence $\delta$.

\subsubsection{Grid Strength and Impedance Effects}
For a grid with Thevenin equivalent $E_{\mathrm{th}}$ and impedance $Z = R + jX$, the voltage angle deviation due to power injection can be approximated as:
\begin{equation}
    \delta \approx \arcsin\left( \frac{P X}{V_1 V_2} \right)
    \label{eq:weak_grid}
\end{equation}
In weak grids (high $X$), small changes in $P$ produce larger $\delta$ shifts.

\subsubsection{Secondary Frequency Restoration}
In microgrids, secondary control restores frequency and corrects long-term angle drift:
\begin{equation}
    \omega_{\mathrm{corr}}(t) = K_i \int_{0}^{t} \left( \omega_{\mathrm{nom}} - \omega(\tau) \right) \, d\tau
    \label{eq:secondary}
\end{equation}
The corrected frequency is then used in (\ref{eq:angle_integration}) to update $\theta_{\mathrm{ref}}$.

\subsubsection{Multi-Unit Load Sharing}
For $N$ grid-forming units with different droop slopes $K_{p,i}$, the steady-state active power sharing ratio is:
\begin{equation}
    \frac{P_i}{P_j} = \frac{K_{p,j}}{K_{p,i}}, \quad \forall i,j \in \{1, \dots, N\}
    \label{eq:power_sharing}
\end{equation}
The resulting $\theta_{\mathrm{ref},i}$ of each unit will differ transiently, but converge to a common frequency in steady state.

\subsection{Droop Control}
The synchronization of static converters by means of power droop follows the operating principle of conventional power plants. The $f-P$ and $V-Q$ droops or the inverse droops were exploited to synchronize multiple converters. The purpose of such droop based methods is to share power among the converters based on power ratings.

\begin{figure}[h]
    \centering
    \includegraphics[width=1\linewidth]{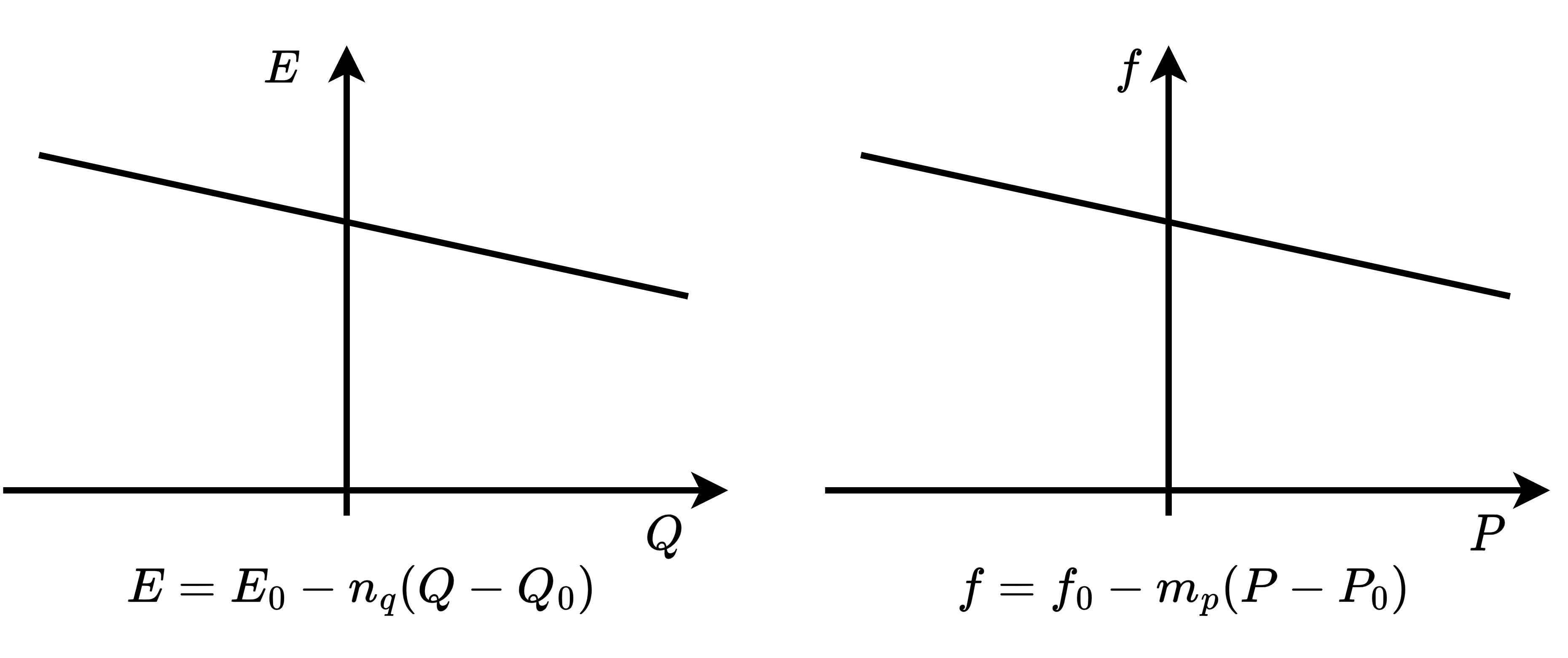}
    \caption{Droop control characteristics}
    \label{fig:enter-droop}
\end{figure}
\begin{figure}[h]
    \centering
    \includegraphics[width=1\linewidth]{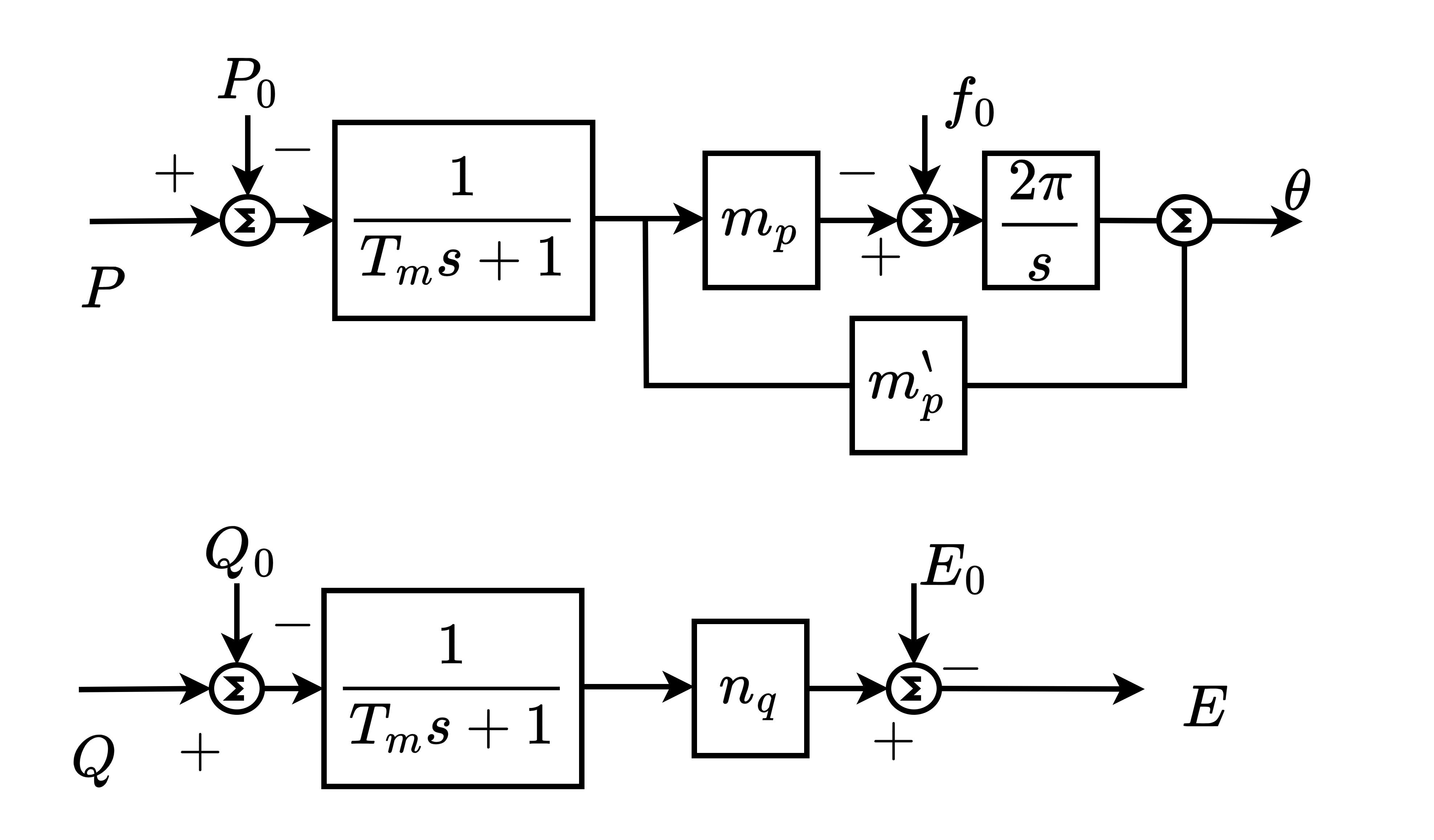}
    \caption{Droop control block diagrams}
    \label{fig:enter-droopc}
\end{figure}
In case of droop control, in most cases, the output power is measured and low-pass filtered with time constants, so the oscillations are eliminated. After the low-pass filter, the output is passed through a gain constant to convert into an internal converter voltage.

\subsection{Loop of Power Synchronisation}
Another way to achieve grid-forming control is through power synchronization control. In this approach, the synchronization mechanism of the swing equation is improved by the researchers in \cite{powersynch}. The capability is improved by including a transfer function which provides damping. The voltage regulation consists of a proportional voltage controller and a reactive power controller, which is optional. As there is a proportional voltage controller, the characteristic is similar to that of $"droop"$. It is assumed that the droop nature considers that the amplitude of voltage at the PCC is linear in nature.
\begin{figure}[h]
    \centering
    \includegraphics[width=0.8\linewidth]{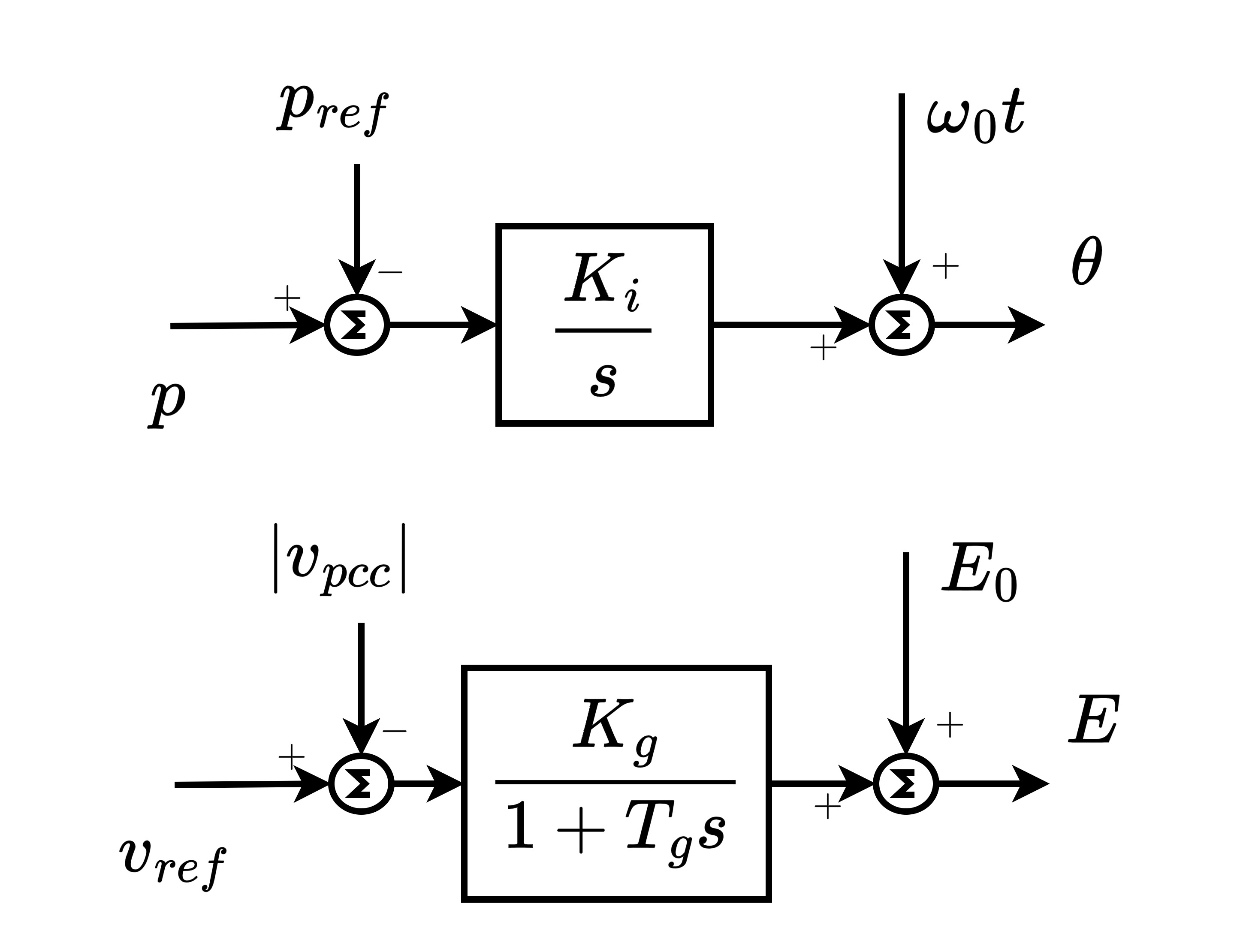}
    \caption{Power synchronisation loop block diagram}
    \label{fig:enter-pdamp}
\end{figure}
\subsection{Virtual Synchronous Machine (VSM)}
In this approach, direct emulation of a synchronous machine is implemented, which is used to generate PWM pulses for the converter to behave as a grid-forming converter. Not all the control approaches succeed in emulating all the dynamics of a synchronous machine, but all of them contain some form of swing equation. Since all the VSM described in \cite{vsm1, vsm2, vsm3} includes swing equation, tha damping of virtual mass inertia is a crucial subject and must be included in the swing damping power term. The swing equation thus becomes as presented below.
\begin{equation}
    J\omega_0\frac{\partial \omega}{\partial t} = P_m-P_e-\frac{D}{2\pi}\omega
\end{equation}
The damping coefficient is an attribute of the synchronous machine and hence remains fixed. However, in the case of VSM control, this parameter is arbitrarily chosen, which is also a reflection of power-frequency droop. As $J$,$\omega_0$, and $2\pi$ are combined to form a single time constant $T_a = 2H$, and a forward damping is introduced in the control block diagrams to damp out oscillations between generators.
\begin{figure}[h]
    \centering
    \includegraphics[width=1\linewidth]{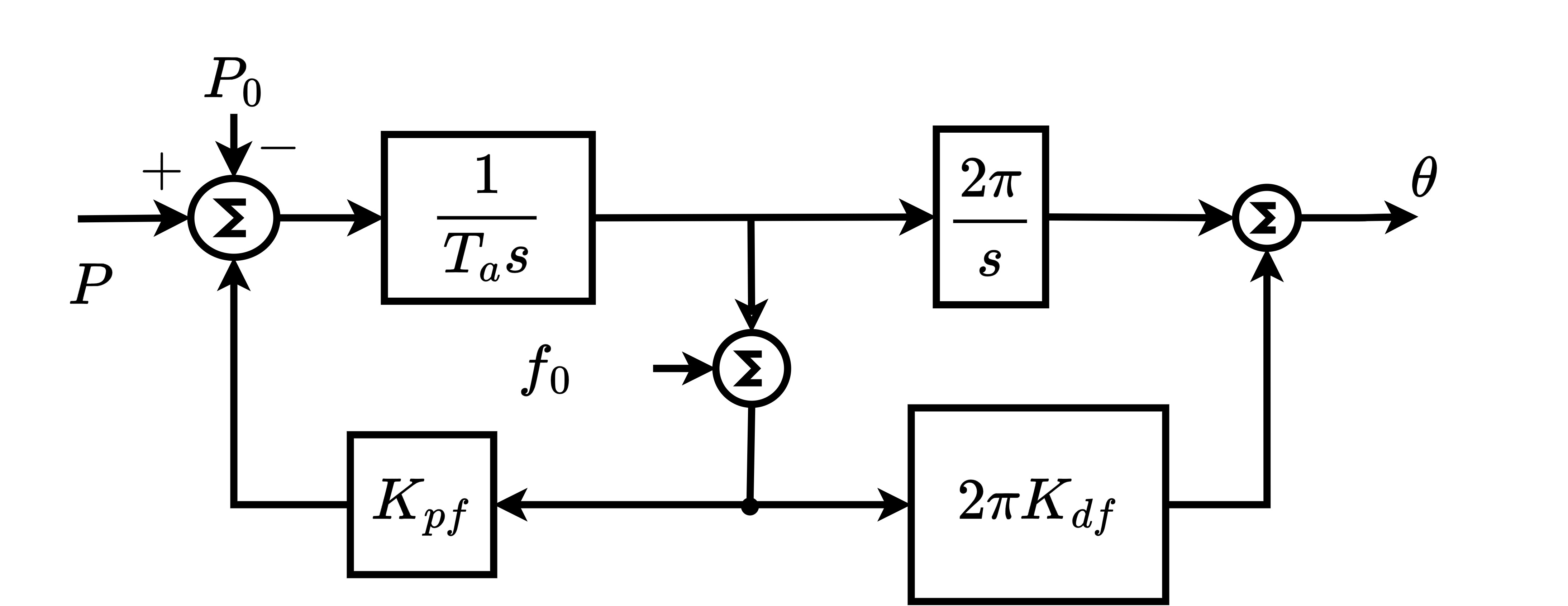}
    \caption{Virtual synchronous machine control block diagrams}
    \label{fig:enter-vsm}
\end{figure}
\begin{figure}[h]
    \centering
    \includegraphics[width=1\linewidth]{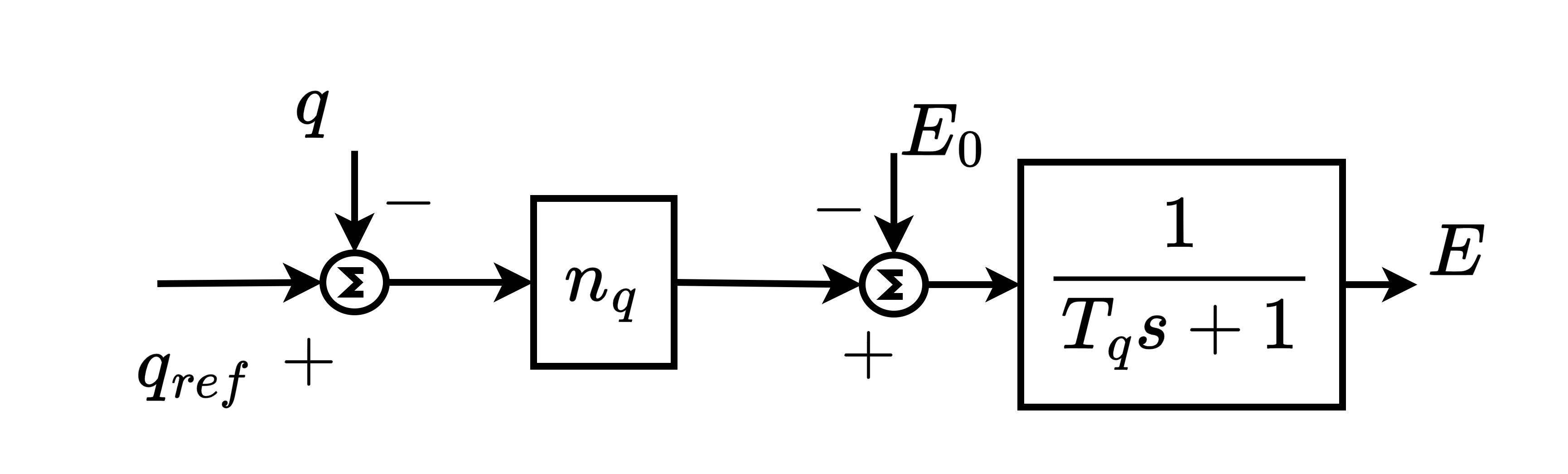}
    \caption{Virtual synchronous machine control block diagrams for reactive droop}
    \label{fig:enter-vsmq}
\end{figure}

The $n_q$ is a slope for Q-V characteristics, and $T_q$ is the excitation time constant for the reactive control loop. The resulting control scheme is shown in the figure.

\subsection{Virtual Oscillator Control (VOC)}
VOC differs from all the aforementioned approaches as it is not based on phasor representation. However, it is a sinusoidal time domain representation which emulates the behaviors as synchronization principles of coupled oscillators \cite{voc1, voc2}
\begin{figure}[h]
    \centering
    \includegraphics[width=1\linewidth]{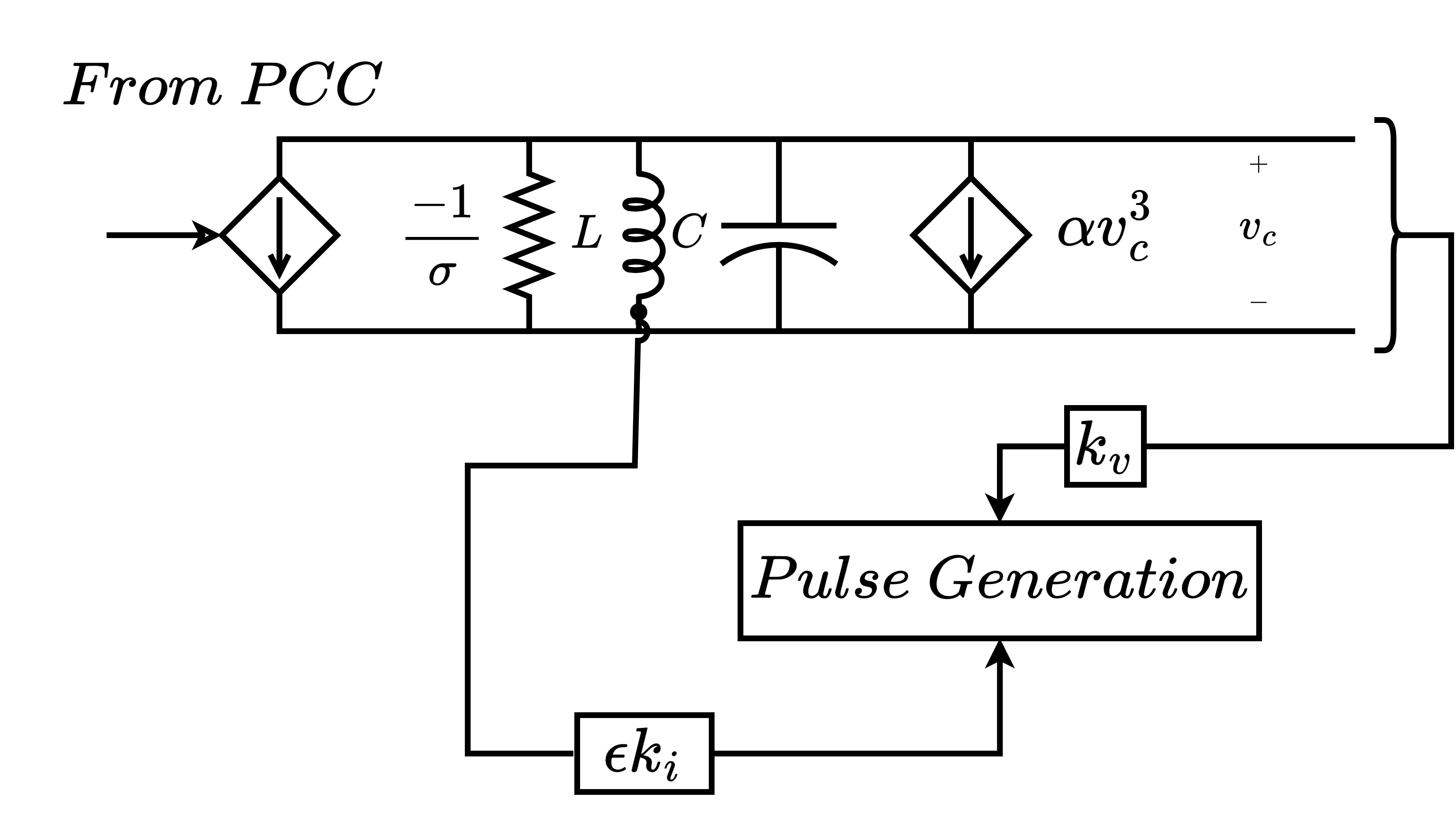}
    \caption{Virtual oscillator control block diagrams}
    \label{fig:enter-vosc}
\end{figure}

The average dynamics is captured by the following equations as the nonlinear dynamics is transferred to the phasor domain.

\begin{equation}
    \dot{V}=\frac{\sigma}{2C}(V-\frac{\beta}{2}V^3)-\frac{k_ik_v}{2CV}Q
\end{equation}
where $\beta$ = $3\alpha(k_v^2 \sigma)^{-1}$

\begin{equation}
    \dot{\theta}=\omega=\omega_0-\frac{k_ik_v}{2CV^2}P
\end{equation}
This equation is subjected to linearisation and change in voltage amplitude, and steady state droop values are determined as explained in \cite{Unruh2020}.

\subsection{Mixed VOC and VSM Control Scheme}
In this type of proposed control scheme, the VOC and VSM contributions are averaged to determine the pulse generation and its effective value is taken as duty cycle to the waveform control.
\subsection{Proposed mixed combination of droop, power synchronisation, voc and vsm control}

In the proposed control scheme the pulse generation and its effective value is apportioned with coefficients and their effective value is determined optimally to varying load conditions formulated from LDC.
\begin{equation}
\theta=\alpha*\theta_{m_p}+\beta*\theta_{vsm}+\gamma*\theta_{psl}+\nu*\theta_{voc}\\
\end{equation}
where
\begin{equation}
     \alpha+\beta+\gamma+\nu = 1
\end{equation}
Few additional constraints are needed to solve the aforementioned problem, which are as shown.
$\alpha \geq 0.8$ ensures droop control remains dominant.
$\beta \leq 0.05$ limits VSM due to possible instability in some grid conditions.
$\nu \leq 0.02$ caps VOC due to its nonlinear time-domain dynamics.

The value of $\alpha$,$\beta$,$\gamma$,$\nu$ is solved using some heuristic algorithm.
$\theta_{m_p}$, $\theta_{vsm}$, $\theta_{psl}$, $\theta_{voc}$ are the corresponding phase angles generated by different control approaches.


\section{Solution to the Mixed control approach for the frequency error minimization problem}
In this paper, a data-driven example of the proposed control scheme is presented, illustrating how to generate the weighted average value of the individual controller juxtaposition. The challenge is to find the proportion of the controller value that remains dominant. In this case, the dictation of the control law based on Mosek solver via nonlinear programming using GAMS remains the main focus of the research. However, a neural network based adaptive controller is designed to solve the proposed problem.
The objective function that is minimized to determine the optimal value of coefficients is shown as below

\begin{equation}
    f_{obj}=\mathbb{MSE}\left((314.159)-(\alpha\theta_{m_p}+\beta*\theta_{vsm}+\gamma*\theta_{psl}+\nu*\theta_{voc})\right)
\end{equation}
\begin{table}[h]
    \centering
    \caption{arbitrary coefficient determination for the aforementioned problem}
    \begin{tabular}{|c|c|c|c|}\hline
        $\alpha$ & $\beta$ & $\gamma$ & $\nu$\\ \hline
        0.974 & 0.009 & 0.012 & 0.006\\ \hline
    \end{tabular}
    \label{tab:my_label}
\end{table}

The problem was formulated using GAMS and the solver MOSEK is called to determine the solution of the problem. However, optimality is yet to be checked. The primal objective function can be observed from the below table.

\begin{table}[h]
    \centering
    \caption{Primal objective values for the minimization problem}
    \begin{tabular}{|c|c|}\hline
        $Iteration\ No.$ & $Primal_{Objective}\ Value$\\ \hline
        1 & 2.384620621e+05\\ \hline
        2 & 1.793568310e+04\\ \hline
        3 & 3.324716586e+03\\ \hline
        4 & 1.754454620e-01\\ \hline
        5 & 6.246100807e-02\\ \hline
        6 & 1.475288364e-02\\ \hline
        7 & 6.133902478e-03\\ \hline
    \end{tabular}
    \label{tab:my_label}
\end{table}

\subsection{ANN based Minimizer}
\begin{figure}[h]
    \centering
    \includegraphics[width=1\linewidth]{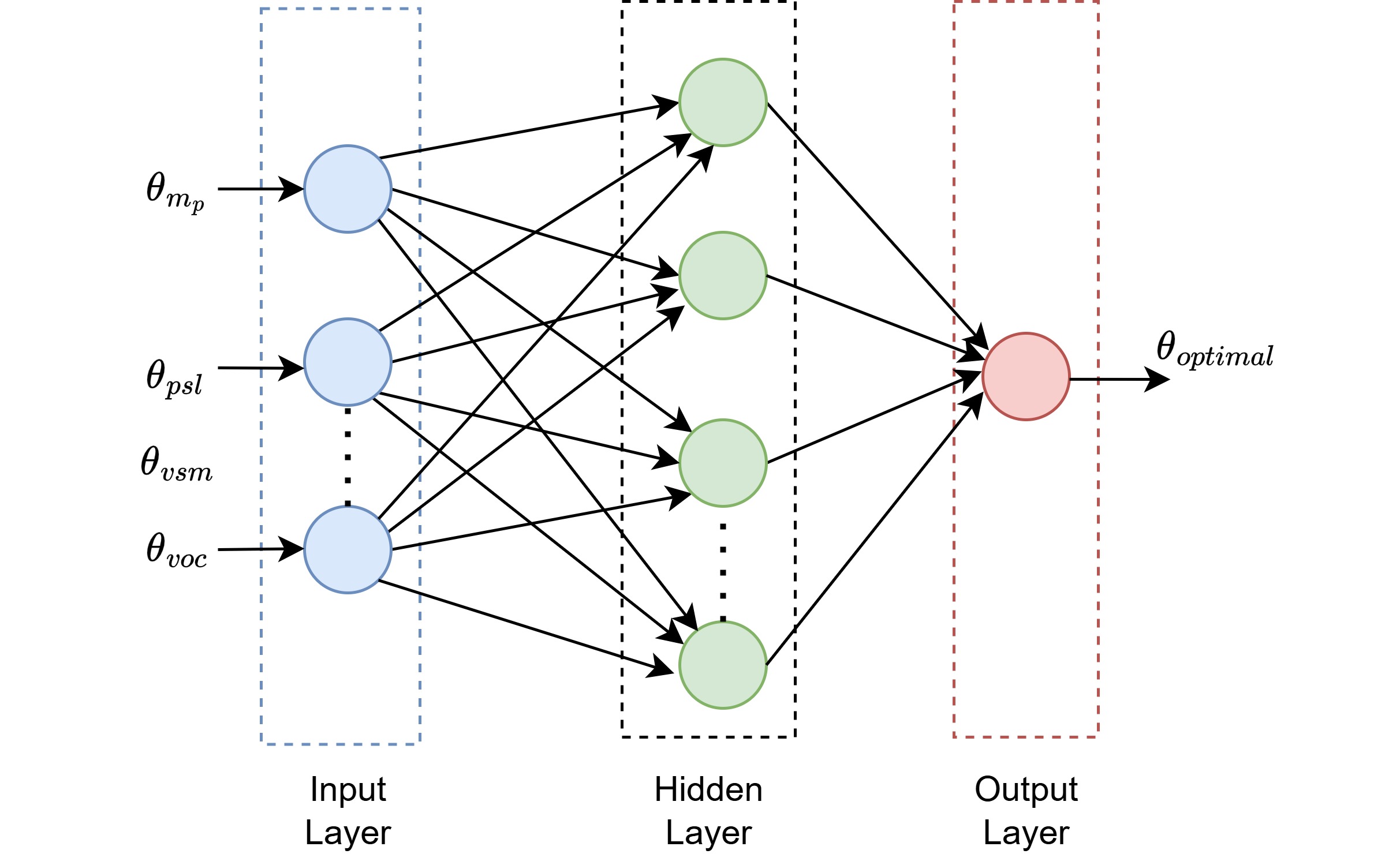}
    \caption{ANN based minimizer}
    \label{fig:ANN}
\end{figure}
In an artificial neural network-based minimizer, the problem is to determine the optimal value of the reference angle subject to frequency error minimization. The cost objective function is dependent on each input. To determine the optimal cost, the first derivative is taken with respect to each input, and the optimal values of each input is determined for test inputs after the network is being trained using training data \cite{deng2025}.

The training samples were generated from a simulation conducted by a large language model using different case scenarios, e.g., overload condition, generation outage, fault scenario, load drop, load surge, microgrid islanding condition, etc. A neural network is trained using a cascade-based backpropagation method using MATLAB. Some results are shown in the figure about the performance of the network. The regression coefficient for all the cases (i.e. training, testing and validation) is 0.99141, which represents the effect sizes, i.e., how much the dependent variable is expected to change when an independent variable changes.

In the case of training of artificial neural network, it is often considered to determine the gradient, $\mu$, and validation check to determine the performance of the artificial neural network.
\paragraph{Gradient}
In ANNs, the gradient is the vector of partial derivatives of the loss function (e.g., MSE, cross-entropy) with respect to the model’s weights and biases. It tells us which direction to adjust parameters to reduce error.
The gradient could be large or small, which defines the weights and updates. In case of a Large gradient, weights are far from optimum, and need bigger updates. In the case of small gradient, the model is closer to optimum, and updates are smaller. 
Usually in practice, gradients are used in backpropagation to propagate error backward. It is also use in optimizers (SGD, Adam, etc.) to update weights.
\paragraph{$\mu$}
$\mu$ usually appears in the Levenberg–Marquardt algorithm, which is a second-order training method used in MATLAB's neural network toolbox.
It acts like a damping factor that balances between the Gauss-Newton method (fast but unstable if $\mu$ is too small), and Gradient descent (slower but more stable when $\mu$ is large).

Training adjusts $\mu$ adaptively in case of step-size reduces, error decrease $\mu$ (trust Gauss-Newton more). If step-size increases, error increases $\mu$ (fall back to gradient descent).
So $\mu$ controls stability vs. speed of learning.

\paragraph{Validation Check}
During training, data is usually split into training, validation, and test sets.
Validation check monitors performance on validation data to prevent overfitting.
In MATLAB (and many ANN frameworks): If validation error keeps increasing for a certain number of epochs (say, 6 checks in a row), training stops early. This is called early stopping.

\begin{figure}[h]
    \centering
    \includegraphics[width=0.95\linewidth]{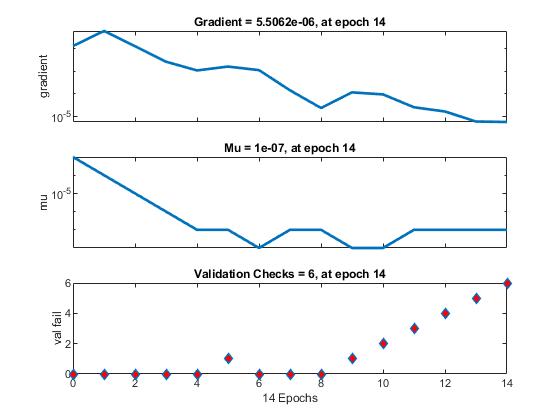}
    \caption{Validation on the Artificial Neural Network for the data samples}
    \label{fig:placeholder}
\end{figure}

\begin{figure}[h]
    \centering
    \includegraphics[width=1.1\linewidth]{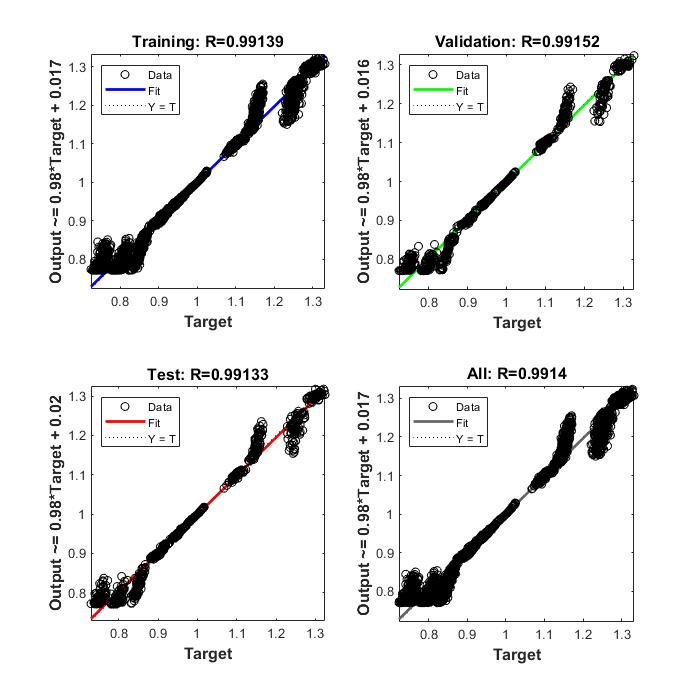}
    \caption{Regression analysis of the trained neural network}
    \label{fig:placeholder}
\end{figure}

The ANN training with the Levenberg-Marquardt algorithm requires a set of inputs and outputs for the determination of weights of the input and hidden layers. In this case, no memory or past inputs were considered for the training purpose, which avoids long-term dependencies. To avoid such problems long short term memory based recurrent neural netowork is used in this condition.

\section{Simulation of Grid forming converter in islanded and grid-connected mode}
In this section, a discussion on grid forming converter in the MATLAB/SIMULINK environment for islanded mode of operation is conducted and the results are explained. The control methodology used in this design is PI-based power synchronization control. Using this control the inverter switches are provided with gate pulses to feed power to the islanded remote load.
\subsection{Parameters for Design of GFM converter}
\begin{figure}
    \centering
    \includegraphics[width=0.8\linewidth]{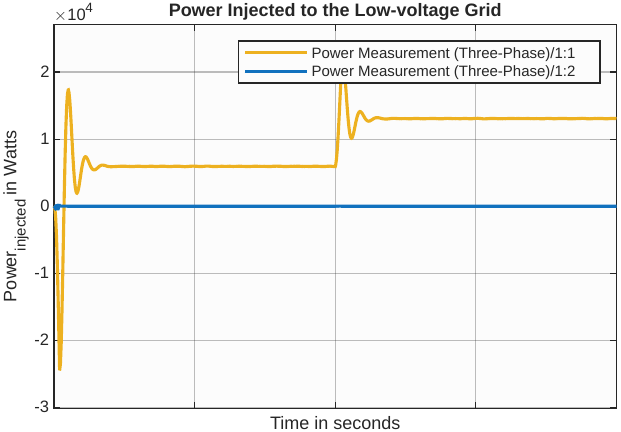}
    \caption{Active power profile for the proposed grid forming control}
    \label{fig:prefgfc}
\end{figure}

\begin{figure}
    \centering
    \includegraphics[width=0.8\linewidth]{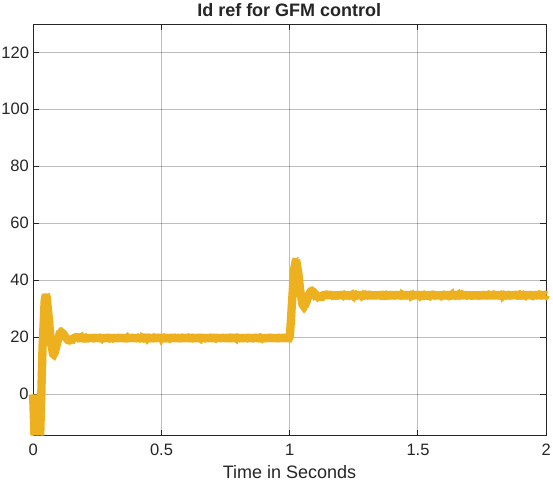}
    \caption{The d-axis current reference for proposed grid forming control}
    \label{fig:idgfc}
\end{figure}

In the design of the GFM converter for the islanded mode of operation, the DC link voltage is taken as 800 Volts. The DC voltage is converted to AC using an inverter with a switching frequency of 20 kHz. The output of the inverter is fed to an LCL filter having an inductance of 0.5 mH and a capacitance of 100 microfarads. The power circuit is fed to a resistive three-phase constant impedance load of 1.1 Ohms with a bus voltage of 480 Volts. The power profile  and $d-axis$ current reference is being depicted for the grid injection in the figure \ref{fig:prefgfc}-\ref{fig:idgfc}.

\begin{figure}
    \centering
    \includegraphics[width=0.95\linewidth]{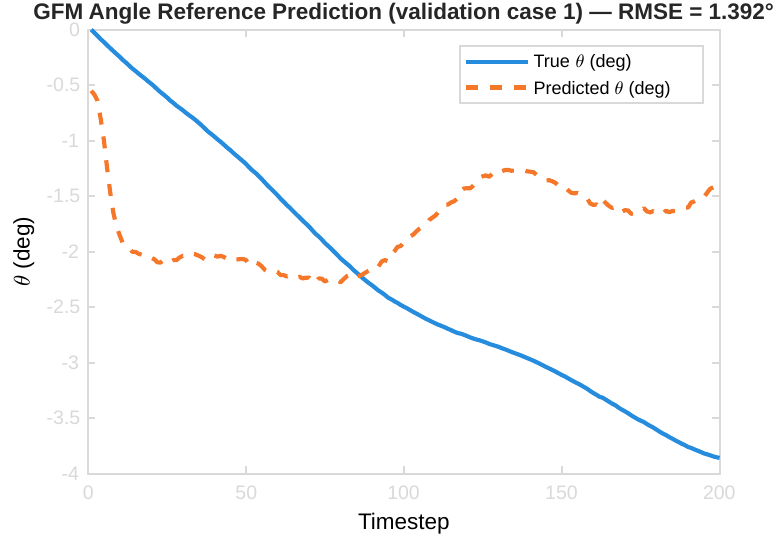}
    \caption{Reference angle generation using LSTM}
    \label{fig:placeholder}
\end{figure}
\begin{figure}
    \centering
    \includegraphics[width=0.9\linewidth]{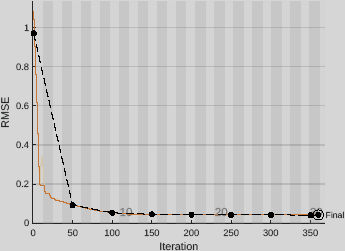}
    \caption{The RMSE of the LSTM Training Performance}
    \label{fig:placeholder}
\end{figure}

\section{Challenges and Future Trends}
In this section, challenges faced by the grid-forming converter during implementation are discussed. The challenges faced are addressed with a suitable solution strategy that is also proposed.
\subsection{Synchronization Stability}
Synchronization with the grid is highly essential for grid-connected converters. In very low short-circuit ratio grids, using a PLL with current control reduces the stability margin of grid-following converters, as converters tend to synchronize with the PCC voltage, which is affected by the output current. Such synchronization introduces negative damping to the utility as claimed in \cite{Rosso2021, dong2015}.
However, in another case, grid-forming converters synchronize with the utility grid based on their output active power using droop control methods like conventional generators. Such power-based synchronization, along with PCC voltage control, accepts the GFM and grid synchronization even in low SCR grids. However, in high SCR stiff grids, a slight phase difference creates large active power oscillations, and the GFM loses synchronism. For wider range of control for SCR conditions, a more robust control should be deployed as shown in paper \cite{harnefors}. 
\par
The inner loop dynamics often interacts with outer voltage loop dynamics, affecting the synchronization stability of the grid-forming converters. This is because of the coupling timescale of operation due to different grid strength conditions or employed inner loop control methods. The effect of inner loop control is discussed comprehensively in \cite{dokus}. In \cite{dVOC2021}, a distributed virtual oscillator control (dVOC) is implemented which investigates transient stability of GFM converter based on Hopf-based oscillator. A comparison is made with conventional droop, whose filter constant is chosen to produce inertia behavior. Even if the critical time is more than CCT, the GFM is able to synchronize after the clearance of the fault in case of dVOC which is not seen in case of droop with virtual inertia.

\subsection{Controller saturation for current limits}
The grid faults initiate the GFM to flow unwarranted current through its static switches, leading to overcurrents that damage the converter circuit. An immediate solution to this is to switch the control mode to vector control mode as proposed in \cite{powersynch}. Using current limitation methods for GFM control, the saturation of PI controllers of cascade control loop, limits the overcurrent. Virtual impedance implementation solves the classical problems such as `$wind-up$' or `$latch-up$', ensuring stability. This approach seems simple in implementation, however, in the case of parallel operation with the synchronous machine, it is rather complex.

\subsection{Fault ride through(FRT)}
In \cite{current_limiting}, reports instability in the outer control loop based on power synchronisation and caused by the limitation of converter current. To resolve such instability, a variable virtual impedance is implemented to increase critical clearing time. However, such an approach does not pan out when the fault is sustained. Another approach to demonstrate GFM behavior during fault and still directly control the converter currents is to limit positive and negative sequence currents as elaborated in \cite{frtfault}.

\subsection{Seamless Mode Transitions}
The transition between islanded and grid-connected operational modes may involve substantial deviations and oscillations, primarily due to potential mismatches in system frequency and voltage amplitude. Grid-forming converter should handle both grid-connected and islanded operation, and transition between them must be smooth, i.e., without any oscillation, which results in instabilities.
To achieve this, droop control is adapted for primary control of the converters and in secondary control, seamless transition operation is included. Mathematically,
\begin{equation}
    \omega = \omega_{ref}+m_p(P-P^*)+\Delta\omega
\end{equation}

\begin{equation}
    V = V_{ref}-n_q(Q-Q^*)+\eta k_I\int (Q-Q^*)dt+\Delta E
\end{equation}
where $\omega$,$E$ converter operated frequency and reference values are taken in subscripts, $m_p$ and $n_q$ represent droop gains, $\Delta\omega$ and $\Delta E$ terms represent compensation terms for secondary control. $\eta$ represents converter operation modes. For design of compensation terms one may refer to \cite{Rosso2021}

\subsection{Critical services sector}
As the energy sector upgrades to include carbon-free generation, it is necessary to understand that the continuity of supply is made to all the sectors of load. The classical way to classify the loads is by their nature of production, i.e., residential, commercial, industrial type \cite{billington}. However, depending on the urgency requirement, the loads are classified into critical and non-critical loads. Critical loads are known as the loads that require immediate attention to their supply during curtailment, whereas non-critical loads can be curtailed during low generation. The examples of critical loads are such sectors that require continuous generation output even during load curtailment, necessitating the requirement of backup generation or reserve generation. Hospitals, universities, etc., come under critical loads. There are converter technologies, such as electric springs \cite{espring}, that facilitate power processing to critical loads during an emergency.


\section{Conclusion}
In this paper, the control methods that can be employed for grid-forming methods are discussed. A mixed control approach to the grid-forming techniques are evaluated using GAMS and performance of the minimization of objective function is discussed. An innovative method is studied for the frequency error minimization problem. Literature suggests gaps in the control algorithms required for GFM based converters which are also discussed in the section VI. The mixed control approach maybe a suitable addition to the research methodology that is being employed by current GFM algorithms. In this paper such a scheme is developed and added to the literature. In the case of MATLAB-based simulation, parallel operation of GFMs is included.

\bibliographystyle{ieeetr}
\bibliography{lib}

\end{document}